\documentclass{preprint}

\usepackage{graphicx}

\usepackage{times}
\usepackage{amsmath}
\usepackage{amssymb}
\usepackage{physics}
\usepackage{color}
\usepackage{upgreek}

\title{Imaging the zigzag Wigner crystal in confinement-tunable quantum wires}

\author{Sheng-Chin Ho,$^{1}$ Heng-Jian Chang,$^{1}$ Chia-Hua Chang,$^{1}$ Shun-Tsung Lo,$^{1}$ Graham Creeth,$^{2}$  Sanjeev Kumar,$^{2}$ Ian Farrer,$^{3, 4}$ David Ritchie,$^{3}$ Jonathan Griffiths,$^{3}$ Geraint Jones,$^{3}$ Michael Pepper,$^{2\ast}$ Tse-Ming Chen$^{1\ast}$}

\begin{document}

\maketitle

\begin{affiliations}
 \item Department of Physics, National Cheng Kung University, Tainan 701, Taiwan
 \item Department of Electronic and Electrical Engineering, University College London, London WC1E 7JE, United Kingdom
 \item Cavendish Laboratory, J J Thomson Avenue, Cambridge CB3 0HE, United Kingdom
 \item Department of Electronic and Electrical Engineering, University of Sheffield, Mappin Street, Sheffield S1 3JD, United Kingdom
\end{affiliations}

\noindent $^\ast$Corresponding author. E-mail:  michael.pepper@ucl.ac.uk (M.P.); tmchen@mail.ncku.edu.tw (T.-M.C.).\\

\begin{abstract}

The existence of Wigner crystallization\cite{wigner_pr34}, one of the most significant hallmarks of strong electron correlations, has to date only been definitively observed in two-dimensional systems. In one-dimensional (1D) quantum wires Wigner crystals correspond to regularly spaced electrons; however, weakening the confinement and allowing the electrons to relax in a second dimension is predicted to lead to the formation of a new ground state constituting a zigzag chain with nontrivial spin phases and properties\cite{meyer_prl07,meyer_jpcm09,welander_prb10,mehta_prl13,klironomos_epl06,klironomos_prb07}. Here we report the observation of such zigzag Wigner crystals by use of on-chip charge and spin detectors employing electron focusing to image the charge density distribution and probe their spin properties. This experiment demonstrates both the structural and spin phase diagrams of the 1D Wigner crystallization. The existence of zigzag spin chains and phases which can be electrically controlled in semiconductor systems may open avenues for experimental studies of Wigner crystals and their technological applications in spintronics and quantum information.

\end{abstract}

Many exotic phases of matter arise when the interactions between electrons become significant, which is more likely to happen in low-dimensional systems due to the reduced charge screening and low, controllable, carrier density. One of the most intriguing examples is the Wigner crystal\cite{wigner_pr34} -- a crystalline phase where electrons are strongly correlated and self-organize themselves in an ordered array to minimize the total potential energy. This situation can arise when the Coulomb interaction dominates over the kinetic energy, and has been clearly observed in two dimensions (2D) for a non-degenerate electron gas on the surface of liquid helium. In confined one-dimensional (1D) systems a trivial Wigner lattice is formed if carriers maintain an identical distance between themselves which will occur in the absence of disorder, and there have been hints for its existence\cite{yamamoto_science06,deshpande_naturephys08}. However, it has been shown both theoretically and experimentally that allowing a 1D electron system to relax in the second dimension would enable observation of the effects of strong interactions which can dominate over the spatial quantization.

Theory\cite{meyer_prl07,meyer_jpcm09,welander_prb10,mehta_prl13} suggests that for quantum wires with tunable confinement potential and carrier density, the strictly 1D Wigner crystal will first evolve into a zigzag Wigner crystal wherein the relative motion of the two coupled chains is locked, and then evlove to form a two-row liquid-like structure as the confinement weakens and/or the carrier density increases. It has also been predicted that the spin-spin interactions in a zigzag Wigner crystal are complicated and can result in a variety of spin phases including a spontaneous spin polarization\cite{meyer_jpcm09,klironomos_epl06,klironomos_prb07}. Such spin chains are not merely of fundamental interests, but also have potential technological applications in spintronics and quantum information, for example, serving as a quantum mediator\cite{antonio_prl15,bose_cp07,awschalom_science13}. So far conductance measurements in quantum wires in which the confinement has been relaxed have shown that a liquid-like two-row structure can be observed as can a conductance jump to $4e^2/h$ that results from the sum of the conductance of each individual row\cite{hew_prl09,gul_jpcm18}. Evidence of coupling between these two rows has also been reported showing an anticrossing of bonding and antibonding states\cite{smith_prb09,kumar_prb14}. However, conductance measurements cannot observe the zigzag Wigner crystal as the dc conductance remains at the quantized value of $2e^2/h$ when measured with leads\cite{maslov_prb95,ponomarenko_prb95}, because there is only one excitation mode for the Wigner crystal as all electrons are tightly locked together. It is therefore highly desirable to develop a technique which can directly image the formation of a zigzag Wigner crystal and assess its consequences.

In order to image the formation of the zigzag Wigner crystals, we create a system in which an on-chip detector is located near a confinement-tunable quantum wire to probe the transverse charge distribution and the spin properties of the wire's electrons as they are emitted. Figures~1a and b illustrate our device structure and the principle of operation. A narrow quantum constriction -- also known as quantum point contact -- is fabricated adjacent to the wire in a GaAs/AlGaAs heterostructures (see Methods) and will be used as a charge detector and spin analyzer. This arrangement composes a magnetic focusing geometry\cite{potok_prl02,folk_science03,chen_prl12,taychatanapat_nphys13,lo_ncomms17}. Electrons injected from the confinement-tunable quantum wire undergo cyclotron motion in the presence of a transverse magnetic field, and are focused towards the charge collector when the cyclotron diameter is equal to the distance $L$ between the injection point and the collector. This gives rise to voltage peaks across the collector (i.e., focusing peaks) at magnetic fields
\vspace*{3mm}\begin{equation}
B = 2\hbar k_{\text{F}}/{eL},
\label{eq1}
\vspace*{3mm}\end{equation} 
where $\hbar$ is the reduced Planck constant, $k_{\text{F}}$ is the Fermi wavevector of the focusing electrons, and $e$ is the elementary charge. Since $L$ is inversely proportional to $B$, the lateral distribution of electrons can now be identified by the magnetic focusing spectrum. As depicted in Fig.~1b, a strictly 1D Wigner crystal (i.e., a line of electrons) will give rise to a single focusing peak -- singlet, with no difference between this and a 1D Fermi liquid, whereas a zigzag structure with two peaks of charge being emitted from different points of the emitter leads to a peak doublet. The difference in magnetic field between the two peaks will give information on their spatial distribution in the emitter.

Figure 1c shows a series of the magnetic focusing spectra as the split-gate voltage $V_{SG}$ is incremented to change the wire confinement, while the top gate is accordingly adjusted so as to keep the wire conductance $G$ and correspondingly, the Fermi energy and carrier density, the same. A clear focusing peak singlet is observed when the quantum wire is strongly confined, indicating that there is only one chain (row). The peak singlet evolves into two peaks which progressively move away from the central point as the confinement is weakened, indicating a transition into two chains (rows). The spatial separation between the two chains can be calculated from the focusing peak separation. It is estimated to be around $200$~nm using equation~1, consistent with the theoretical predictions\cite{welander_prb10}. We also note that the corresponding conductance traces of the wire do not show a jump to $4e^2/h$ (Supplementary Fig.~1) as can occur for a crossing\cite{hew_prl09,kumar_prb14}, which implies that there could be an anti-crossing as discussed\cite{smith_prb09,kumar_prb14} or the system is in the zigzag configuration in which a single chain is distorted to two locked chains within a single wavefunction\cite{meyer_prl07,meyer_jpcm09}.

This technique allows investigation of the self-arrangement of the electrons with increasing density at a fixed confinement. Figure~2a demonstrates that with increasing $G$ (which corresponds to increasing Fermi energy and electron density) a peak singlet suddenly changes to a peak doublet at $G \sim 80$~$\mu$S. The observed transition from one to two chains of electrons with increasing $G$ is in agreement with previous theoretical predictions that a strictly 1D phase will change to a zigzag Wigner crystal phase once the Coulomb repulsion dominates over the confinement potential. Figure~2b shows a series of focusing spectra as in Fig.~2a but for an extremely shallow confinement potential in the wire. The peak doublet appears all the way from $G = 10$ to $150$~$\mu$S since the confinement potential is much weaker compared to Fig.~2a, thereby Coulomb repulsion always dominates.

To have a better understanding of the competition between electron density and the confinement potential of the wire and consequently how the interplay between them leads to different phases, we map the characteristics of the focusing peaks as a function of wire conductance and split-gate voltage. Peaks with well-defined doublet (singlet) structures are indicated by yellow (red), peaks with ambiguous shapes are represented in orange. The corresponding phase diagram showing how the strictly 1D and the two-chain structures inter-relate is plotted in Fig.~2c. Remarkably, this reveals a dome-shaped phase diagram, which qualitatively agrees with the theoretical expectations\cite{meyer_prl07,mehta_prl13} for the interaction-driven zigzag Wigner phase transition.

The zigzag Wigner crystal is expected to have rich and intricate spin properties. The ring-exchange interactions among three, four, and larger numbers of spins plays a significant role in such zigzag structures, which is distinct from the strictly 1D Wigner crystal in which only the nearest-neighbor exchange coupling occurs. Ring exchanges involving an odd number of spins are well known to favor ferromagnetism, and hence both fully and partially spin polarized phases were theoretically predicted\cite{klironomos_prb07,meyer_jpcm09} to appear in zigzag Wigner crystals depending on the competition between different ring-exchange couplings. It is worth noting that such ring-exchange induced spin polarization should be a gapless spin excitation\cite{meyer_jpcm09} and cannot be inferred from the wire conductance. Hence the spin properties in such quasi-1D Wigner crystals remain experimentally unexplored unless the spin content can be probed in a direct way.

We now demonstrate that the charge collector can be tuned to further act as a spin detector (or filter) such that the spin properties of the electrons within the quantum wire can be directly probed. This is done by turning the collector into a spin selective (i.e. spin polarized) mode allowing the parallel spins to pass while the antiparallel spins cannot. Therefore, the voltage drop across the collector (i.e., the focusing peak height) depends on the spin polarization of both the electron being injected ($P_\text{inj}$) and the collector ($P_\text{c}$), and can be described by\cite{potok_prl02,folk_science03,chen_prl12}
\vspace{3mm}\begin{equation}
 V_{c}=\alpha {\frac{h}{2e^{2}}}I_\text{inj}(1+P_\text{inj}P_\text{c}),
\vspace{3mm}\end{equation}
where $I_\text{inj}$ is the current being injected and $\alpha$ is a spin-independent parameter accounting for the loss of current during the focusing process. This method has been shown to be extremely useful in assessing the spin polarization in various quantum systems simply by measuring the collector voltage\cite{potok_prl02,folk_science03,chen_prl12,lo_ncomms17,chuang_nnano15}.

Here we apply a large source-drain DC bias to bring the collector into a fully spin polarized phase known as the $0.25 \times 2e^2/h$ structure\cite{chen_prl12,chen_apl08} (Supplementary Fig.~2). In other words, $P_\text{c}$ can be modulated from $0$ to $1$ as the source-drain DC bias current $I_{DC}$ across the collector is increased. Figure 3a shows how the peak doublet and singlet vary with increasing $I_{DC}$, when the collector conductance is fixed at $0.2 \times 2e^2/h$, i.e., below the 0.25 structure to guarantee the occurrence of spin polarization once $I_{DC}$ increases\cite{chen_prl12,chen_apl08}. The height of the focusing peak doublet (top panel of Fig.~3a) rises significantly as the source-drain DC bias is increased from $I_{DC}=0$, but begins to saturate when $I_{DC} \geq 25$~nA. Note that the saturation of the peak height indicates that $P_c$ has approached its maximum value of $1$ when $I_{DC} \geq 25$~nA, which is also consistent with the observation of the fully spin polarized 0.25 structure. The spin polarization of the two chains of electrons is, therefore, estimated to be $\sim 60 \%$ ($P_\text{inj} \approx 0.6$). The fact that the peak doublet is spin polarized furthermore implies that the two chains of electrons within the wire have formed a zigzag Wigner crystal phase since the spin polarization is anticipated only in the zigzag phase\cite{klironomos_prb07,meyer_jpcm09}. In contrast, the peak singlet (bottom panel of Fig.~3a) barely changes as $I_{DC}$ increases from $0$ to $40$~nA. This suggests that $P_\text{inj} \approx 0$ and therefore the spin is unpolarized, as expected for a strictly 1D quantum wire comprising a single row which is not distorted by the interactions.

Figure 3b shows the spin phase diagram of our confinement-tunable quantum wire as a function of conductance and split-gate voltage, similar to Fig.~2c. The spin polarization $P_\text{inj}$ is estimated using the ratio of the focusing peak height when the collector acts as a spin filter (i.e., $P_\text{c} \approx 1$ at $I_{DC} = 20$~nA) to that when it acts just as a charge collector (i.e., $P_\text{c} \approx 0$ at $I_{DC} = 0$~nA). Therefore, the peak ratio is approximately equal to $1+P_\text{inj}$. Here, in order to double check the validity of the spin phase diagram, an alternative analysis method to assess $P_\text{inj}$ is applied and gives consistent results (Supplementary Fig.~3). We notice that the spin polarization $P_\text{inj}$ is in general higher as the confinement widens, and is nearly $100\%$ when the wire potential is extremely shallow and the conductance (and hence electron density) is large enough for the interactions to be sufficiently strong, consistent with the appearance of the two chains reported in Fig.~1 \& 2 as well as the theoretical predications\cite{meyer_prl07,klironomos_prb07,meyer_jpcm09}. The consistency can be clearly seen in a direct comparison of the spin phase diagram (Fig.~3b) with the structural phase diagram of the Wigner crystallization (Fig.~2c). Furthermore, we note that the spin polarization $P_\text{inj}$ is lower and could be down to nearly $0\%$ when the peak doublet is less well-defined and, accordingly, the separation between the two chains is smaller. The structural and spin phase diagrams provide access and new insights into the 1D zigzag Wigner crystal, and could be the foundation for future theoretical studies of the electron-electron and spin interactions.

We now focus on the temperature dependence of the focusing peaks to characterize the melting of the zigzag Wigner crystal. The temperature-dependent evolution of the peak doublets, the peaks which cannot be unambiguously identified as singlet or doublet, and the peak singlets are measured from $T = 0.03$~K to $4.5$~K, as shown in Fig.~4a, b, and c, respectively. All types of focusing peaks are observed to fall with increasing temperature due to thermal smearing\cite{taychatanapat_nphys13}. Interestingly, we find the peak doublet (which represents the zigzag) evolves into a single peak (a single chain) when the temperature rises above $\sim 3$~K, indicating that the zigzag Wigner crystal has collapsed. A similar evolution is also observed for the peak with the ambiguous shape (Fig.~4b). For comparison, the focusing peak singlet (Fig.~4c) maintains its form and position with increasing temperature up to $4.5$~K.

To further assess the melting temperature of the zigzag Wigner crystal, we also study how the degree of the spin polarization -- a property which is linked to the zigzag Wigner crystal phase -- varies with increasing temperature. Figure~4d plots the spin polarization $P_\text{inj}$ of the three different types of focusing peaks corresponding to those presented in Fig.~4a-c as a function of temperature. $P_\text{inj}$ for the two chains (black squares) remains around $0.95$ at $T\lesssim 1$~K but begins to fall when the temperature is increased beyond $\sim 1$~K, and eventually reaches $0$ when $T\gtrsim 3$~K. Similar behavior is found where the doublet structure is not so well-defined (red circles), albeit with smaller $P_\text{inj}$ at $T\lesssim 1$~K. The spin polarization for the strongly confined case (blue triangles) is $P_\text{inj} \approx 0$ regardless of temperature variation even though the peak height itself reduces significantly with increasing temperature as shown in Fig.~4c. We note that the temperature associated with the melting of the 1D zigzag Wigner crystal reported here is about one order of magnitude higher than those of the 2D Wigner crystal\cite{chen_nphys06,deng_prl16,jang_nphys17}, but close to the temperature associated with the Tomonaga-Luttinger Liquid in 1D systems\cite{jompol_science09,laroche_science14}.

The experiments presented here do not merely demonstrate the 1D zigzag Wigner crystallization but, more importantly, provide fundamental new insights into its true nature (including the spin phases and properties, the separation between the two zigzag chains, and the intimate connection between these two things), as well as how the Wigner crystal develops with respect to confinement, Fermi energy, and temperature. These properties were largely inaccessible in past experiments. Hence, our results and the method of using an on-chip charge and spin detector employing electron focusing to study 1D nanowires should inspire more theoretical and experimental works in both fundamental and technological aspects. Wigner spin chains with a variety of phases that can be electrically controlled in solid-state systems appear promising as a building block for realizing novel spintronics and quantum circuits.

\section*{Method}

The devices are fabricated on a GaAs/AlGaAs heterostructure grown using molecular beam epitaxy. The layer structure in reverse order of growth is as follows: $10$~nm GaAs cap, $200$~nm Si-doped Al$_{0.33}$Ga$_{0.67}$As, $75$~nm undoped Al$_{0.33}$Ga$_{0.67}$As spacer, and $1$~$\mu$m GaAs buffer. The two-dimensional electron gas (2DEG) formed underneath the wafer surface has an electron carrier density and mobility of $1.1 \times 10^{11}$~cm$^{-2}$ and $4 \times 10^{6}$~cm$^{2}$V$^{-1}$s$^{-1}$, respectively. The mean free path is around $22$~$\mu$m, much longer than the cyclotron trajectories in our magnetic focusing devices. Our device schematic is shown in Fig.~1a \& b. The device consists of a confinement-tunable quantum wire on the left, formed by a pair of split gates of width $W = 1$~$\mu$m and length $L = 0.4$~$\mu$m, and a $0.6$~$\mu$m-long top gate separated by a $45$~nm thick SiO$_{2}$ insulator. The narrow quantum constriction on the right, which acts as both a charge collector and spin analyzer, is $0.4$~$\mu$m wide and $0.2$ $\mu$m long.

The experiments are performed in a dilution refrigerator at a base temperature of $20$~mK unless otherwise stated. In order to simultaneously measure the quantum wire conductance, the collector conductance, and the magnetic focusing spectrum, we use two independent excitation sources of (i) a $37$~Hz AC voltage of 50~$\mu$V to the quantum wire and (ii) a $77$~Hz AC current of 1~$n$A plus a DC current $I_{\text{DC}}$ to the collector. The focusing signal (i.e., the voltage developed across the collector) is measured at a reference frequency of $37$~Hz to detect the focusing electrons traveling from the quantum wire to the collector.


\section*{Acknowledgements}
We thank L. W. Smith and C.-T. Liang for discussions. This work was supported by the Ministry of Science and Technology (Taiwan), the Headquarters of University Advancement at the National Cheng Kung University, and the Engineering and Physical Sciences Research Council (UK).

\begin{figure}
\begin{center}
\includegraphics[width=1\columnwidth]{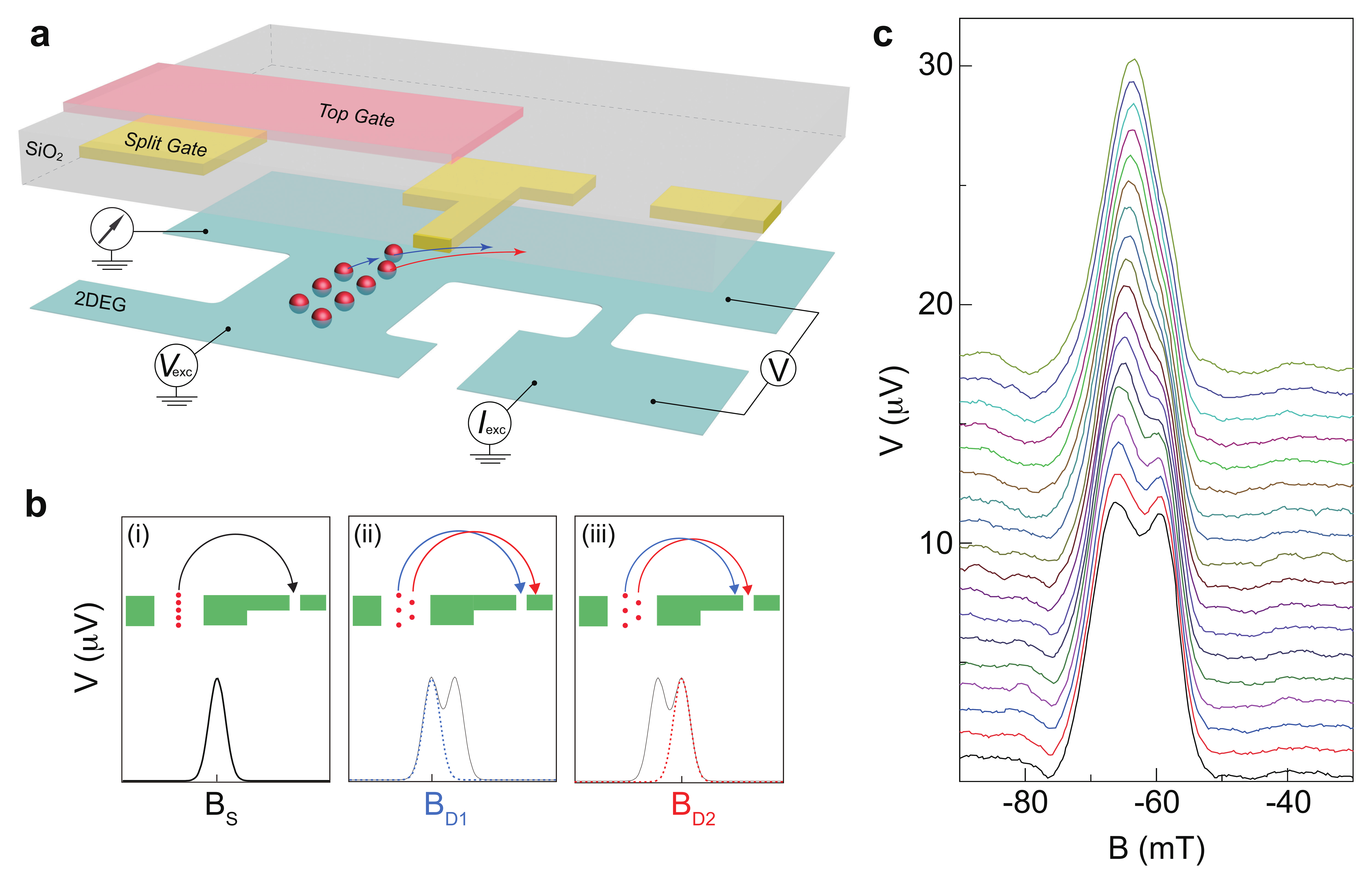}
\end{center}
\caption{\small \textbf{Imaging the zigzag Wigner crystallization.} \textbf{a}, Schematic of our device to image the formation of zigzag Wigner crystal. \textbf{b}, Illustration of the magnetic focusing peaks and their corresponding cyclotron motions. A single chain of electrons will give rise to a single peak as illustrated in (i) when focused into the detector, whereas electrons in a double chain are focused into the detector at two different magnetic fields, as illustrated in (ii) and (iii), giving rise to a peak doublet. \textbf{c}, Magnetic focusing spectrum at various split gate voltages from $-2.7$ to $-1$~V (from top to bottom) in steps of $0.1$~V while the top gate voltage is accordingly varied to fix the wire conductance at $100$~$\mu$S.}
\end{figure}

\begin{figure}
\begin{center}
\includegraphics[width=1\columnwidth]{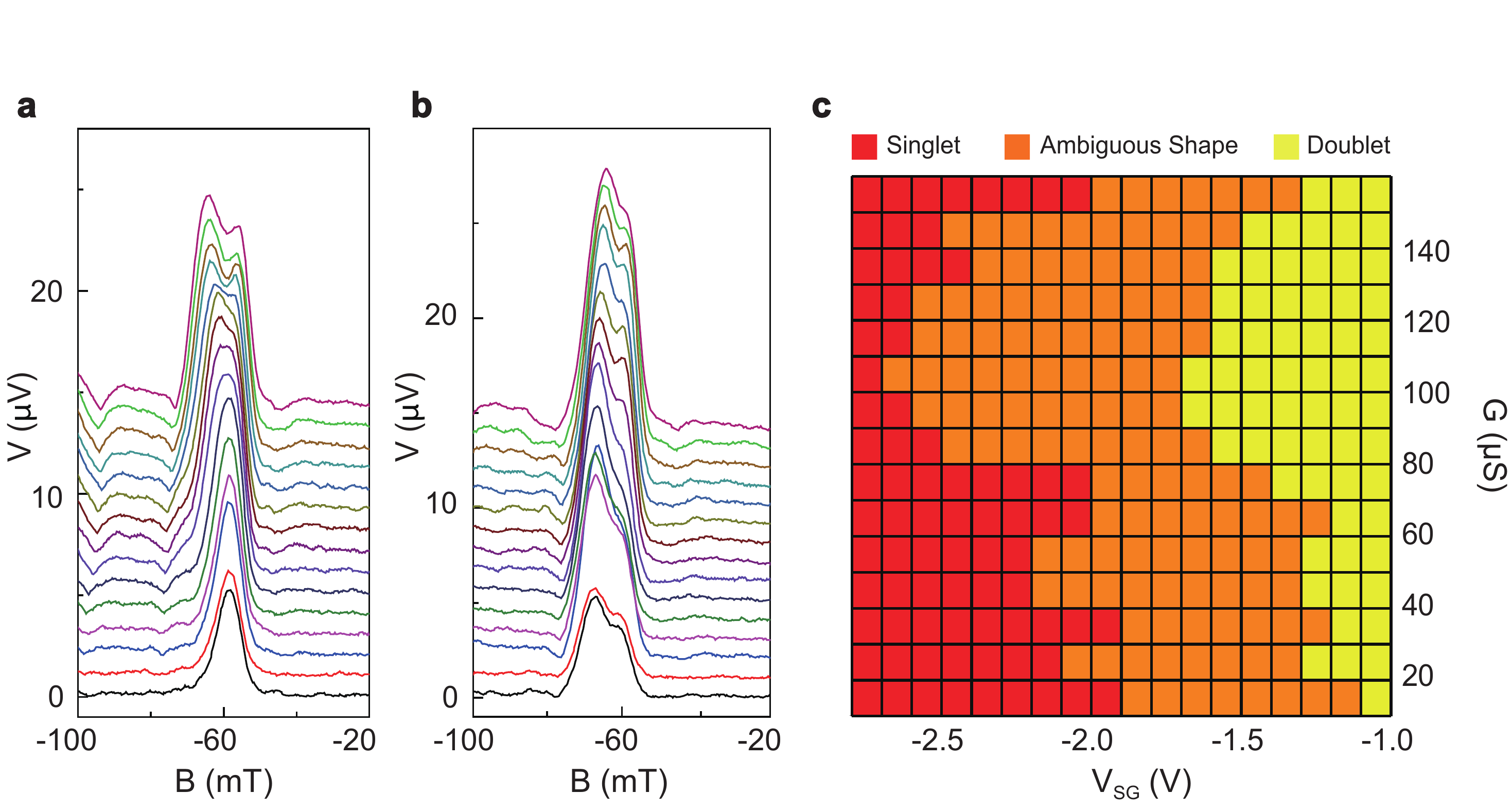}
\end{center}
\caption{\small \textbf{Gate controlled Wigner crystallization.} \textbf{a,b}, The magnetic focusing spectra plotted as a function of magnetic field at various conductance values from $10$ to $150$~$\mu$S when the split gate voltage is fixed at $V_{SG}= -2$~V (\textbf{a}) and at $V_{SG} = -1.4$~V (\textbf{b}). \textbf{c}, A map of the characteristics of the magnetic focusing peaks as a function of conductance and split gate voltage settings. Data obtained from a different cooldown. Well-defined doublet (singlet) peak are mapped in yellow (red) colour while the peaks with ambiguous shapes are represented in orange.}
\end{figure}

\begin{figure}
\begin{center}
\includegraphics[width=1\columnwidth]{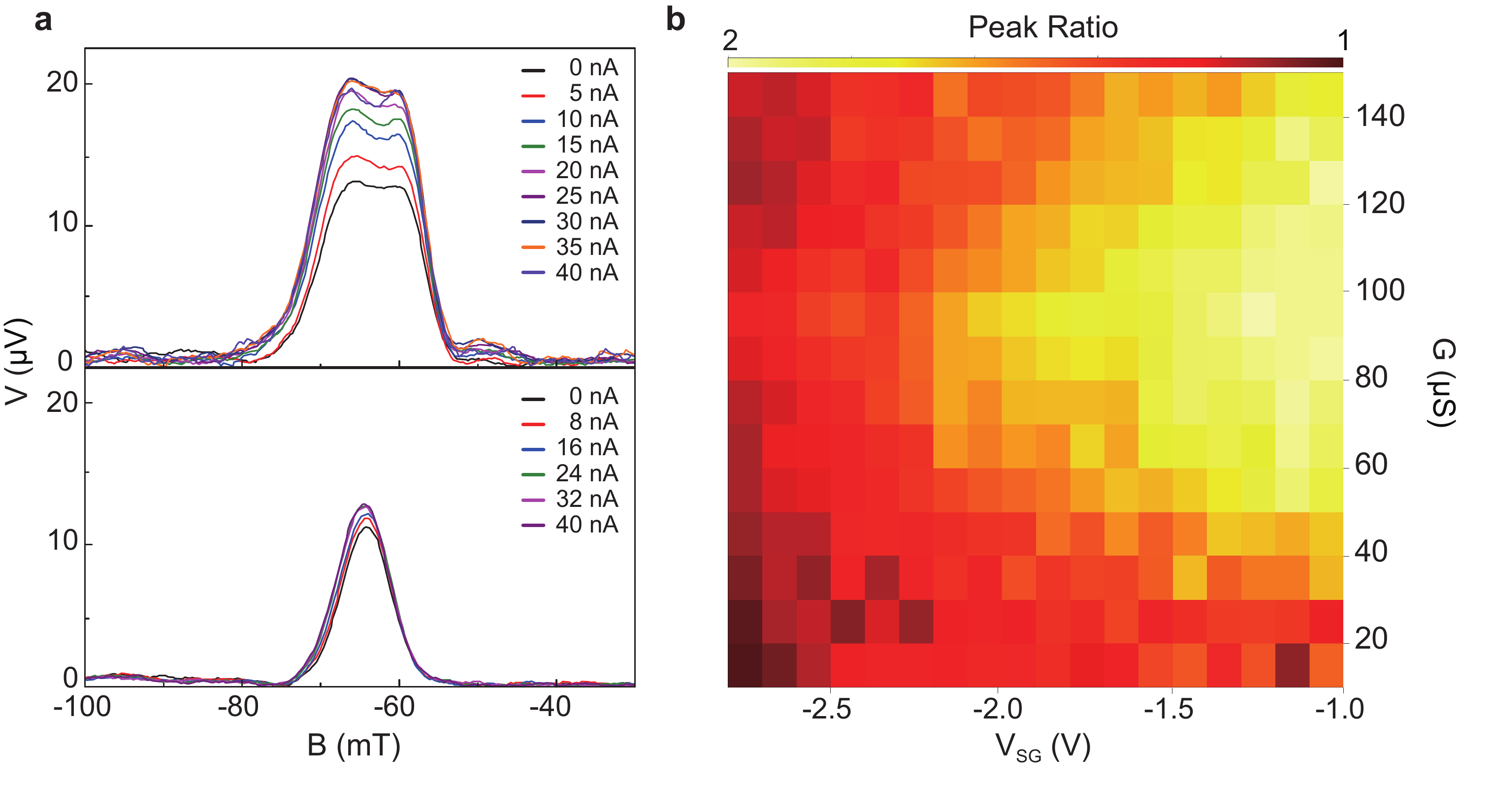}
\end{center}
\caption{\small \textbf{Spin properties of the Wigner crystal.} \textbf{a}, The magnetic focusing peak doublet (top panel) and singlet (bottom panel) at various collector dc source-drain biases from $0$ to $40$~nA, where the collector conductance is fixed at $0.2 \times 2e^2/h$. \textbf{b}, The spin phase diagram mapped out using the ratio of the magnetic focusing peak when the QPC collector is tuned as a spin detector to that when it is tuned as a charge detector, plotted as a function of conductance and split gate voltage settings. 
}
\end{figure}

\begin{figure}
\begin{center}
\includegraphics[width=1\columnwidth]{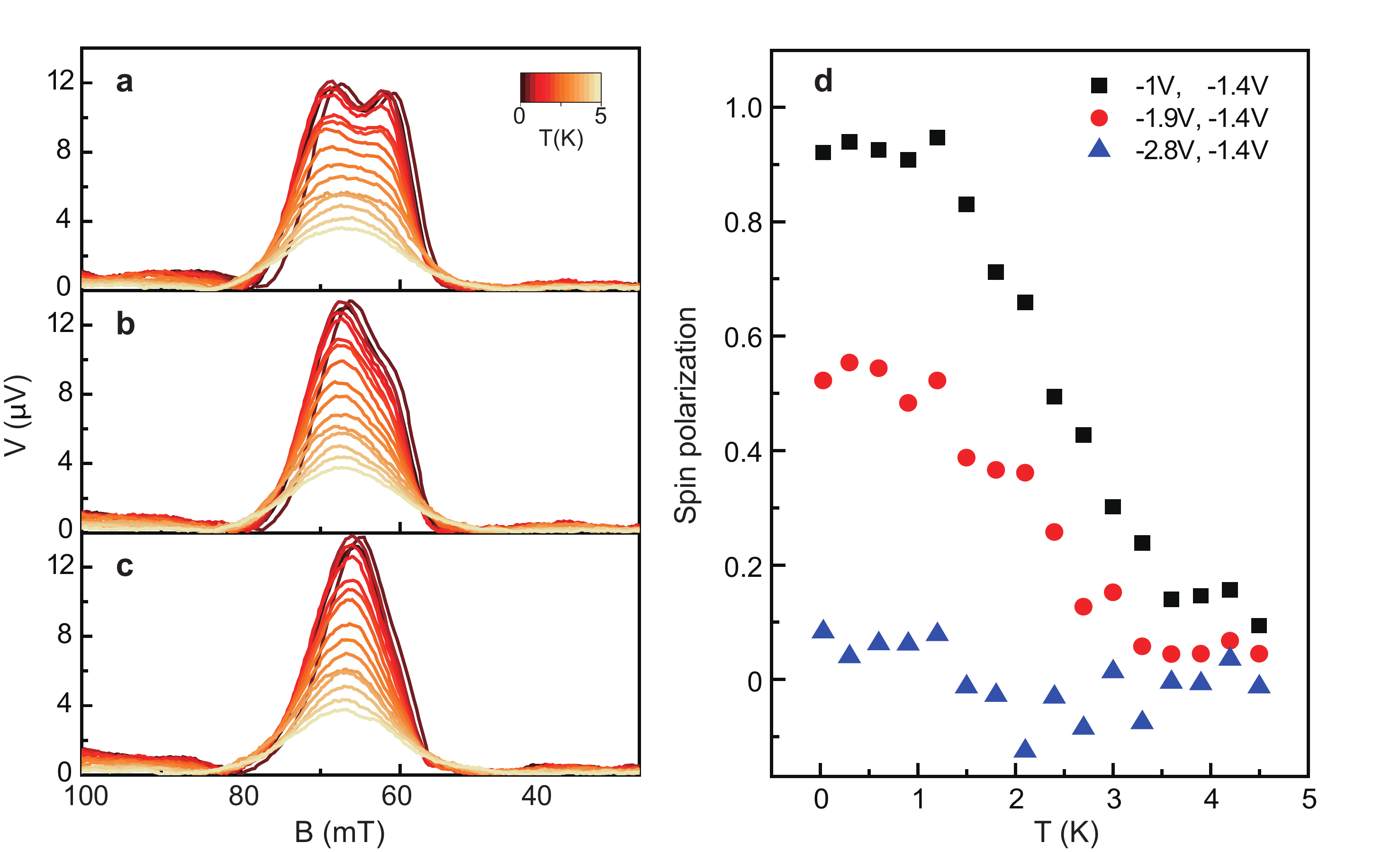}
\end{center}
\caption{\small \textbf{Temperature dependence of the zigzag Wigner crystallization and their spin properties.} \textbf{a}-\textbf{c}, Evolution of the magnetic focusing peak with increasing temperature for focusing peak doublet (\textbf{a}), peaks without a well-defined shape (\textbf{b}), and peak singlet (\textbf{c}). \textbf{d}, The spin polarizations, which are estimated using the same method as in Fig.~3b, as a function of temperature for the three focusing peaks shown in \textbf{a}-\textbf{c}. 
}
\end{figure}

\end{document}